# Engineering Dzyaloshinskii-Moriya interaction in B20 thin film chiral magnets

Emrah Turgut[1], Hanjong Paik[2], Kayla Nguyen[1], David A. Muller[1,3], Darrell G. Schlom[2,3], and Gregory D. Fuchs[1,3]

1. School of Applied and Engineering Physics, Cornell University, Ithaca, NY 14853, USA
2. Department of Materials Science and Engineering, Cornell University, Ithaca, NY 14853, USA
3. Kavli Institute at Cornell for Nanoscale Science, Ithaca, NY 14853, USA

Chiral magnetic $Mn_xFe_{1-x}Ge$ compounds have an antisymmetric exchange interaction that is tunable with the manganese stoichiometric fraction, $x$. Although millimeter-scale, polycrystalline bulk samples of this family of compounds have been produced, thin-film versions of these materials will be necessary for devices. In this study, we demonstrate the growth of epitaxial $Mn_xFe_{1-x}Ge$ thin films on Si (111) substrates with a pure B20 crystal structure in the stoichiometric fraction range $x$ from 0 to 0.81. Following systematic physical and magnetic characterization including microwave absorption spectroscopy, we quantify the antisymmetric exchange interaction and helical period as a function of $x$, which ranges from 200 nm to 8 nm. Our results demonstrate an approach to engineering the size of magnetic skyrmions in epitaxial films that are grown using scalable techniques.

I. **Introduction**

Controlling exchange interactions in noncollinear magnets not only allow an exploration of the microscopic phenomenology of noncollinear magnetism, but it also enables the engineering of chiral spin textures, including topological magnetic skyrmions [1,2]. One of these magnetic interactions is the antisymmetric exchange interaction, known as the Dzyaloshinskii-Moriya interaction (DMI), which occurs when inversion symmetry is broken at interfaces or in the volume of non-centrosymmetric materials and generates twisted spin alignments [1,3–5]. A well-known group of these materials is the B20 transition metal monosilices and monogermanides, e.g., FeGe, MnSi, and $Fe_xCo_{1-x}Si$ [2].

Among B20 compounds, isostructural solid solutions $Mn_xFe_{1-x}Ge$ are special because they alone show chiral magnetism through the entire compositional range $x$ (from 0 to 1), with a corresponding helical period ranging from 200 nm to 3 nm [6,7]. Because skyrmion diameters closely match to the helical period in B20 materials, controlling composition also means controlling magnetic skyrmion size, which is particularly important for achieving high density in memory and logic devices based on magnetic skyrmions. Moreover, $Mn_xFe_{1-x}Ge$ displays remarkable physical properties, including multiple phase transitions between short-range and long-range orderings and a transition between the DMI and the effective Ruderman-Kittel-Kasuya-Yosida exchange interaction [8,9]. Experimental studies of $Mn_xFe_{1-x}Ge$ compounds have, however, been limited to millimeter-sized polycrystalline bulk samples due to the requirement of high-pressure synthesis [7,10] to achieve the desired metastable B20 polymorph, which has prevented the full exploration and exploitation of these novel properties relevant to technological applications. In this regard, the use of epitaxy to stabilize and grow thin films [11] of the desired B20 polymorph on a technologically relevant substrate like silicon will enable the scalable fabrication of magnetic devices that can take advantage of non-collinear spin textures.

In this work, we demonstrate the epitaxial growth of $Mn_xFe_{1-x}Ge$ thin films using molecular-beam epitaxy (MBE) on a Si (111) substrate. We find that we can grow this material with a pure B20 crystal structure in the wide compositional range of $x$ from 0 to 0.81. Using high-resolution scanning transmission electron microscopy (STEM), we validate the epitaxy of $Mn_xFe_{1-x}Ge$, and using X-ray diffraction (XRD), we quantify the in-plane and out-of-plane lattice strain as a function of manganese fraction, $x$. Then, following conventional magnetic characterization, we perform spinwave spectroscopy and corresponding micromagnetic simulations. Combining the results of these measurements, we establish the critical temperature, the strength of the symmetric/antisymmetric exchange interactions, and the helical periods in our films as a function of $x$. Our work demonstrates an approach to controlling the helical period through composition over a range of 200 nm to 8 nm, thus enabling the engineering of magnetic skyrmion diameter in scalably grown thin-film chiral magnets.

## II. Growth & physical characterization

We grow $Mn_xFe_{1-x}Ge$ thin films by co-depositing from Mn, Fe, and Ge sources onto the surface of a B20 FeSi seed layer at 200 °C using a Veeco GEN10 MBE system with a $2 \times 10^{-9}$ Torr base pressure. The FeSi seed layer is created by the room-temperature deposition of a monolayer of iron on a 7x7 reconstructed Si (111) surface followed by flash annealing at 500 °C. The formation of B20 $Mn_xFe_{1-x}Ge$ is continuously monitored by *in-situ* reflection high-energy electron diffraction (RHEED) to ensure the right stoichiometric B20 phase of $Mn_xFe_{1-x}Ge$ without segregation (See supplementary material for RHEED patterns). The deposition rate of the growth is approximately 5 Å/min and two hours of growth produces a 60-nm thick film of each composition, which is determined by low-angle X-ray reflectivity (XRR) measurements.

Next, to confirm an epitaxial relation between Mn$_x$Fe$_{1-x}$Ge and the Si (111) substrate, we image a cross-section of Mn$_{0.18}$Fe$_{0.82}$Ge film using high-resolution STEM at a 300 keV electron beam energy (Fig. 1(a)). We find that the FeSi seed layer is approximately 1 nm thick where we clearly observe a transition from the silicon diamond crystal structure to the B20 crystal structure. The Mn$_x$Fe$_{1-x}$Ge epitaxial layer also shows high crystalline quality as observed by the two parallelograms of B20 cubic crystal structure [12].

Strain plays a key role in establishing the magneto-static anisotropy, and thus the stable magnetic phases in thin-film B20 compounds [13–16]. To calculate the film strain we measure $\theta$-$2\theta$ XRD scans using a Rigaku Smartlab X-ray Diffractometer. In particular, to determine the out-of-plane and in-plane strains, we measure the out-of-plane lattice spacing along [111] ($d_{111}$) at $\chi = 90°$ and the lattice spacing along [100] ($d_{100}$) at $\chi = 35.26°$ [17]. For each film, we calibrate the instrument against the known lattice constant of the silicon substrate, 5.431 Å, at both $\chi$ angles [18]. To find the reference bulk lattice constants for Mn$_x$Fe$_{1-x}$Ge, we first perform energy-dispersive X-ray spectroscopy characterization of each film to obtain the Mn:Fe ratio precisely. Then, using Vegard's law [19,20] and the known lattice constants of FeGe and MnGe [6,21,22], we calculate the reference bulk lattice constants. By comparing the measured lattice spacings to the reference values, we find the in-plane and out-of-plane strains in the films as shown in Fig. 1(b).

Although the lattice constants of our Mn$_x$Fe$_{1-x}$Ge films are larger than the corresponding lattice spacing of the silicon substrate (5.431√3/2=4.703 Å) for $x > 0.17$ our films consistently have lateral expansion and perpendicular compression, regardless of $x$. This suggests that the strain deformation in the cubic B20 crystalline thin films is most likely due to the mismatch in the thermal expansion coefficient between the film and Si, rather than the direct room temperature lattice mismatch [23]. Nevertheless, we find that by growing Mn$_x$Fe$_{1-x}$Ge thin films using MBE, we are able to reduce the strain by a factor of four as compared to the strain in sputtered B20 FeGe thin films [16]. This is a crucial step towards reducing the unfavorable easy-plane anisotropy in B20 thin films [13–15].

### III. Magnetic Characterization

After confirming the high-quality epitaxial growth of our films, we perform systematic magnetic characterization using a vibrating sample magnetometer (VSM). Before we explain the magnetometry measurements, we first describe the magnetic behavior of B20 chiral magnet thin films. Because B20 chiral thin films have easy-plane anisotropy [14,16,24,25], we can assume the uniformity of magnetization in the plane and define the one dimensional free energy density as

$$\mathcal{H} = A(\partial_z \boldsymbol{m})^2 - \boldsymbol{D}.\boldsymbol{m} \times \partial_z \boldsymbol{m} - \boldsymbol{H}.\boldsymbol{m} - K_u(\boldsymbol{m}.\hat{\boldsymbol{n}})^2 - \frac{1}{2}\boldsymbol{H_m}.\boldsymbol{m}, \qquad (1)$$

where $A$ is the exchange stiffness constant, $D$ is the DMI constant, $H$ is the external magnetic field, $K_u$ is the anisotropy constant, and $H_m$ is the demagnetizing field due to the shape of the sample [26]. This chiral magnetic free energy presents two important parameters: the zero field helical wave vector $Q=2\pi/L_D=D/2A$, where $L_D$ is the helical wavelength, and the in-plane saturation field $H_D = D^2/(2AM_s)$, where $M_s$ is the saturation magnetization [26]. Previous studies have shown that these two parameters, the helical period and the saturation field, determine many properties in chiral magnets, including spin resonance dynamics [16,27]. Additional studies have found that bulk $Mn_xFe_{1-x}Ge$ samples have substantial variations in the values of $Q$ and $H_D$ as a function of the stoichiometric fraction $x$ [7,8,28]. For example, $Q$ for FeGe is 0.09 nm$^{-1}$ and it approaches zero at $x_c = 0.2$, which is the critical fraction at which the DMI constant goes through zero and changes its sign. For $x > x_c$, $Q$ starts to increase again, eventually reaching 2 nm$^{-1}$, corresponding to a helical period, $L_D = 3$ nm for $x = 1$ (MnGe) [8,9]. As we mentioned above, $H_D$ is also proportional to $D^2$, varying from a few mT to 13 T through the whole range of $x$ [9]. Due to variations of $Q$ and $H_D$ from the previous observations, it is important to extract these quantities in our $Mn_xFe_{1-x}Ge$ thin films using magnetometry.

To find the saturation field and magnetization of our films, we perform magnetometry measurements using a Quantum Design VSM with an in-plane applied field. To make consistent comparisons between the films, we measure the saturation fields at 40 K for all of the films. This is an ideal temperature for comparison because at higher temperatures material with large $x$ show strong fluctuations in the helical phase [10], while at lower temperatures the saturation field reaches our instrument maximum. In Fig. 2(a), we show the magnetization vs. in-plane magnetic field plots for $x = 0.08, 0.18, 0.26$, and $0.34$, and in Fig. 2(b), we show the plots for $x = 0.44, 0.56, 0.62, 0.72$ and $0.81$. By finding the minimum in the second derivative of $M$ with respect to $H$ ($\partial^2 M/\partial^2 H$, see supplemental material), we find the saturation field as a function of $x$, which we plot with pink squares in Fig. 2(c). Although $H_D$ is less than 100 Oe for the films near the $x_c = 0.2$ [7,8], the maximum $H_D$ is found at 38 kOe for the $Mn_{0.81}Fe_{0.19}Ge$ film, the film with the highest manganese fraction in the series. This behavior of the saturation field agrees perfectly with the variation in DMI as a function of $x$ in polycrystalline bulk $Mn_xFe_{1-x}Ge$ alloys [8–10]. We also find the saturation magnetization, $M_s$, from the $M$ vs. $H$ measurements, plotted with blue triangles in Fig 2(c). In contrast to the dramatic variation in $H_d$, we observed a slight, approximately 35 %, increase in $M_s$ as a function of $x$, which is also consistent with previous observations in bulk crystals [9].

To find the critical temperature ($T_c$) of the paramagnetism-to-helimagnetism transition and the spinwave stiffness constant ($D_{sw}$) for the films, we perform additional magnetometry measurements as a function of temperature. To find $T_c$ for each film, we apply a small magnetic field in the plane (160 Oe for $x = 0.08, 0.18, 0.26$, and $0.34$, and 320 Oe for the remaining manganese fractions) and measure the magnetization as the temperature is varied between 300 K and 100 K. Then, we differentiate the magnetization with respect to $T$; this gives the susceptibility of the films ($\partial M/\partial T$) and indicates the helical phase boundaries [29], which we use to find the maximum of the curve for each film to determine $T_c$ (Fig 2(d)). We find that the lowest six manganese stoichiometric fractions display a similar trend. The $\Delta M/\Delta T$ curves for $x = 0.62$, $0.72$, and $0.81$ show large fluctuations below $T_c$, which was also observed with bulk $Mn_xFe_{1-x}Ge$ measurements due to the helical phase fluctuations [10,30]. In contrast to the bulk materials [28], we observe an increase in $T_c$ for the films with high manganese fraction, i.e., $T_c$ saturates to around 200 K for $x > 0.5$ (Fig. 2(e)). In this range, the value of $T_c$ is approximately 50 K higher than in bulk versions of $Mn_xFe_{1-x}Ge$ [28]. We speculate that geometrical confinement and surface twist effects can change the energy landscape (the last two right-hand-side terms in Eq. (1)) for the helimagnetism against the thermal fluctuations, which could increase $T_c$ in our films [26].

To find $D_{sw}$ as a function of $x$, which can be used to obtain the $A$ and DMI constants, we measure the temperature-dependent $M_s$ by applying a saturating magnetic field. At low temperatures, well below $T_c$, the magnetization follows the Bloch-$T^{3/2}$ law (See supplemental material for validation), and thus $D_{sw}$ can be extracted by fitting to

$$M_s(T) = M_s(T=0)\left(1 - \frac{g\mu_B\eta}{M_s(T=0)}\left(\frac{k_BT}{D_{sw}}\right)^{3/2}\right), \quad (2)$$

where $g$ is the electron's gyromagnetic ratio (taken to be $g = 2$ [32,33]), $\mu_B$ is the Bohr magneton, $k_B$ is the Boltzmann constant, $\eta$ is a dimensionless geometrical factor of magneton density that can be approximated as 0.0587 for films thicker than 50 nm [32,34]. Then, we measure the magnetization of our films by varying the temperature between 80 K and 150 K. Because the silicon substrates produce a large diamagnetic background in addition to the films magnetizations, there are significant error bars in $D_{sw}$, as shown in Fig. 2(e). We note that $D_{sw}$ follows $T_c$, which is expected because of the Bloch-$T^{3/2}$ law [28,32].

Using the relations $D_{sw} = \frac{g_e\mu_BH_D}{Q^2}\left(1 + \frac{1}{2}Q^2a_0^2\right)$ and $H_D = D^2/(2AM_s)$, where $a_0$ is the lattice constant [16,35–37] (See supplemental material for details), we find the helical wave vector $Q$, the exchange stiffness $A$, and the DMI coefficient $D$ in $Mn_xFe_{1-x}Ge$ thin films (Table I). In addition, by using $L_D = 4\pi A/D$, we calculate the helical period $L_D$ and plot it as a function of $x$ in Fig. 3(a). We note that the helical period of the film with the highest manganese fraction is 7.6 nm, which is the shortest helical period in a thin film reported to date. A 7.6 nm helical period should correspond to a 9 nm ($2\times7.6/\sqrt{3}$) skyrmion lattice constant, which is yet to be experimentally confirmed in these films. There has been some epitaxial growth of B20 MnGe thin films [38,39], which should contain smaller helical periods, around 4 nm [6], however, their magnetic structures have not been reported. Finally, we note that the physical parameters in Table I are crucial for micromagnetic simulations of the helimagnetic spin dynamics that are discussed in the next section.

IV.  **Spinwave resonances: Experiments and simulations**

An important method to characterize ferromagnetic materials and quantify their magnetic properties, e.g., $M_s$ and anisotropy fields, is microwave absorption spectroscopy (MAS) [40]. This is particularly true for helical magnets because the helical magnetic structure has unique spin resonances that depend strongly on the helical period and the saturation magnetic field [16,33]. Therefore, we exploit MAS to unambiguously confirm the helical spin texture and the strength of the DMI in $Mn_xFe_{1-x}Ge$ thin films.

Before we perform MAS measurements, we first computationally study spin dynamics and calculate the resonance frequencies of the helical phases using micromagnetic simulations with Mumax3 software [34]. In these simulations, we first find the equilibrium spin texture for a given value of $x$, $H$, and $T$, and then extract the natural oscillation frequencies that are triggered by a magnetic field impulse. See Ref. [16] for further simulation details.

To illustrate the helical wrapping of the equilibrium spin textures in $Mn_xFe_{1-x}Ge$ thin films, we plot the simulated x-component of the magnetization ($M_x$) as a function of film thickness at $H = 0$ Oe field (Fig. 3(b)). We have used a +1 vertical shift between curves for clarity. For $x = 0.08$, 0.18, 0.26, and 0.34, less than a full helical period is supported within the 60 nm film thickness. For larger values of $x$, multiple helical wrappings are accommodated due to the strong DMI.

Next we computationally find the spinwave profiles as a function of thickness and frequency, which are sensitively related to the helical alignment [16]. Samples with $x \leq 0.26$ do not have helical resonances because they do not have a full wrapping period within the film thickness, however, they do have Kittel thickness mode spinwaves [See supplemental material]. Samples with high manganese fraction ($x = 0.44–0.81$) show higher resonance frequencies than the films with fractional wrapping for the same number of nodes. In Figs. 4(a) and 4(b), we plot the simulation results for three stoichiometric fractions, $x = 0.44$, 0.56, and 0.62, because their resonance frequencies are well-matched to our experimental apparatus. In particular, by spatially averaging the Fourier transform of the natural oscillations, we find the total amplitudes of the resonances as a function of frequency, Figs. 4(a1)–4(a3). The film with $x = 0.44$ has three resonance frequencies ($f_1 = 3.7$ GHz, $f_2 = 7.6$ GHz, and $f_3 = 12.2$ GHz) that have one, two, and three antinodes in spinwave modes (Figs. 4(a1) and 4(b1)), respectively. The film with $x = 0.56$ has two resonances ($f_1 = 5.1$ GHz and $f_2 = 10.5$ GHz) with one and two antinodes (Figs. 4(a2) and 4(b2)). Similarly, the film with $x = 0.62$ has two resonances ($f_1 = 6.4$ GHz and $f_2 = 13.2$ GHz) with one and two antinodes (Figs. 4(a3) and 4(b3)).

To experimentally examine the spin dynamics in our $Mn_xFe_{1-x}Ge$ thin films, we place each sample film-side down on a broadband coplanar waveguide (CPW) [42]. The CPW is located inside a cryostat that is placed between the poles of an electromagnet that can reach a DC magnetic field of ±7 kOe. In addition, we apply a modulated magnetic field with 6 Oe amplitude and 219 Hz frequency. We demodulate the microwave power transmitted through the CPW with respect to the magnetic field modulation using a lock-in amplifier, enabling sensitive, low-noise MAS measurements. See Refs. [16,43] for further experimental details. Next, we perform MAS on the films with $x = 0.44$, 0.56, and 0.62 in which we vary the microwave frequency and the DC magnetic field at 40 K. In Fig. 4(c), we plot the MAS voltage as a function of the RF frequency between 2 and 14 GHz and the magnetic field between -5 and 5 kOe.

To compare with the $H = 0$ Oe simulation results, we focus on the experimental absorption profiles near zero magnetic field. In particular, the film with $x = 0.44$ shows three resonances at 3.6, 6.8, and 12.4 GHz (Fig. 4(c1)), which is in close correspondence with the numerically predicted spin wave resonances shown in Fig. 4(a1). The film with $x = 0.56$ has $f_1 = 5$ GHz and $f_3 = 10$ GHz, which are also in close agreement with the simulation, however, the experimentally-observed resonance at $f_2 = 8.5$ GHz did not appear in the simulation, perhaps because it is broad in comparison to the other resonant features. For the $x = 0.62$ film shown in Fig. 4(c3), we experimentally observe a resonance at $f_1 = 5$ GHz, which is lower than the numerically predicted resonance. Furthermore, we experimentally observe an extremely complicated resonant feature at 13.5 GHz. Although a resonance at this frequency is predicted at $H = 0$ Oe by micromagnetic simulation (Fig. 4(b3)), the complexity of the lineshape initially led us to discount this feature as an artifact. Nevertheless, it appears across multiple sample pieces with the same composition, multiple CPWs, and even a different signal generator. Therefore, we conclude that it is a real resonant absorption process by the film, but one that requires additional study. We also note that the MAS signal for the $x = 0.62$ sample is one order of magnitude smaller than the MAS signal for the $x = 0.44$ sample, and the thus measurement is more susceptible to noise, which could lead such a disordered MAS signal at 13.5 GHz.

V.   **Discussion and Conclusion**

Understanding the spinwave dynamics in DMI-varied $Mn_xFe_{1-x}Ge$ thin films could play a crucial role in future applications like frequency tunable spin-torque oscillators using chiral magnets [44,45]. Although we grow this material series using MBE, the growth could be adapted to more scalable approaches such as magnetron sputtering based on the growth conditions presented here. Furthermore, even though our films have helical spin texture below room temperature, it is a good model system for understanding behavior and device integration of thin-film materials with crystalline broken inversion symmetry leading to a strong DMI. Such strong DMI can exhibit sub-ten nm magnetic skyrmions, which is essential for high-density memory and spintronic applications [46].

Mn$_x$Fe$_{1-x}$Ge thin films are also a good model system for future transport studies of chiral magnets. The detailed understanding of Berry phase in chiral magnets remains a complicated problem due to the broken inversion symmetries in both momentum and real space of electrons in Mn$_x$Fe$_{1-x}$Ge compounds. There have been very few studies that tackle this problem theoretically and experimentally using polycrystalline, mm-size materials [9,47–49], and DMI varied Mn$_x$Fe$_{1-x}$Ge thin films will be useful to address this problem by allowing arbitrary size device fabrication. Moreover, we speculate that volume DMI materials like Mn$_x$Fe$_{1-x}$Ge thin films will have advantages relative to interfacial DMI thin-film heterostructures due to more a uniform skyrmion diameter. In recent studies of Pt/Co/Ir multilayers, which have interfacial DMI [50,51], the anomalous Hall effect is used to detect skyrmion motion under a charge current and large fluctuations in the Hall signal were observed due to the varying diameter of the magnetic skyrmions. In this respect, Mn$_x$Fe$_{1-x}$Ge thin films are desirable for a uniform skyrmion diameter and a small variation in Hall signals. In addition, real-time imaging of skyrmion motion under a charge current and sensing the resulting Hall voltage in bulk DMI materials is a challenge at the forefront of the field of chiral magnetism. This is partially because of the limited understanding and availability of B20 thin films for complicated device fabrication [50,51]. Our effort here is an important step to overcome this challenge.

In conclusion, we present the high-quality epitaxial growth of Mn$_x$Fe$_{1-x}$Ge thin films on a silicon substrate within the compositional range of $x$ from 0 to 0.81. We perform comprehensive physical and magnetic characterization. We find that the films have a positive strain in the plane and a negative strain out of the plane for all values of $x$, which suggests that the thermal expansion mismatch plays a key role in establishing film strain. Nonetheless, we are able to reduce the strain in B20 thin films by a factor of four as compared to sputtered FeGe films, which is promising for reducing the uniaxial anisotropy in chiral thin films. Through magnetometry studies, we quantified the saturation field, magnetization, exchange stiffness and DMI strength as a function of $x$. Using the powerful combination of MAS measurements and micromagnetic simulations, we validate the presence of helical spin textures through their unique spin wave resonances in three Mn$_x$Fe$_{1-x}$Ge ($x$ = 0.44, 0.56, and 0.62) thin films. Our results demonstrate an approach to controlling the helical pitch and magnetic skyrmion diameter through compositional control, which supports the integration of thin-film chiral magnets into future high-density, power-efficient spintronic devices.

### Acknowledgments

This work was supported by the DOE Office of Science (Grant No. # DE-SC0012245). MBE growth was carried out using the thin film growth and characterization facilities of the Platform for the Accelerated Realization, Analysis, and Discovery of Interface Materials (PARADIM), a National Science Foundation Materials Innovation Platform (DMR-1539918). We also acknowledge use of facilities of the Cornell Center for Materials Research (CCMR), an NSF MRSEC (Grant No. # DMR- 1719875). We further acknowledge facility use at the Cornell Nanoscale Science and Technology Facility (Grant No. # ECCS-1542081), a node of the NSF-supported National Nanotechnology Coordinated Infrastructure.

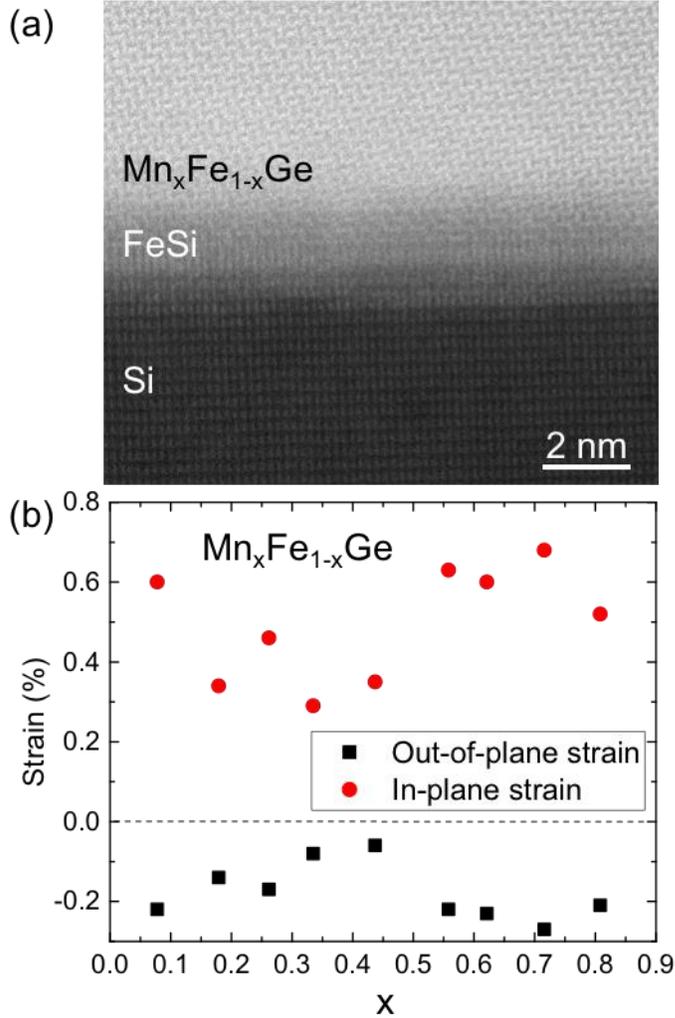

**FIG. 1.** Physical characterization of $Mn_xFe_{1-x}Ge$ thin films. (a) A high-resolution cross-sectional STEM image of thin-film $Mn_{0.18}Fe_{0.82}Ge$ along the $<1\bar{1}0>$ zone axis that shows epitaxial growth on top of the FeSi seed layer. (b) In-plane and out-of-plane strains in $Mn_xFe_{1-x}Ge$ thin films plotted as a function of the manganese stoichiometric fraction $x$, which shows compression in the direction of the film normal and expansion in the plane.

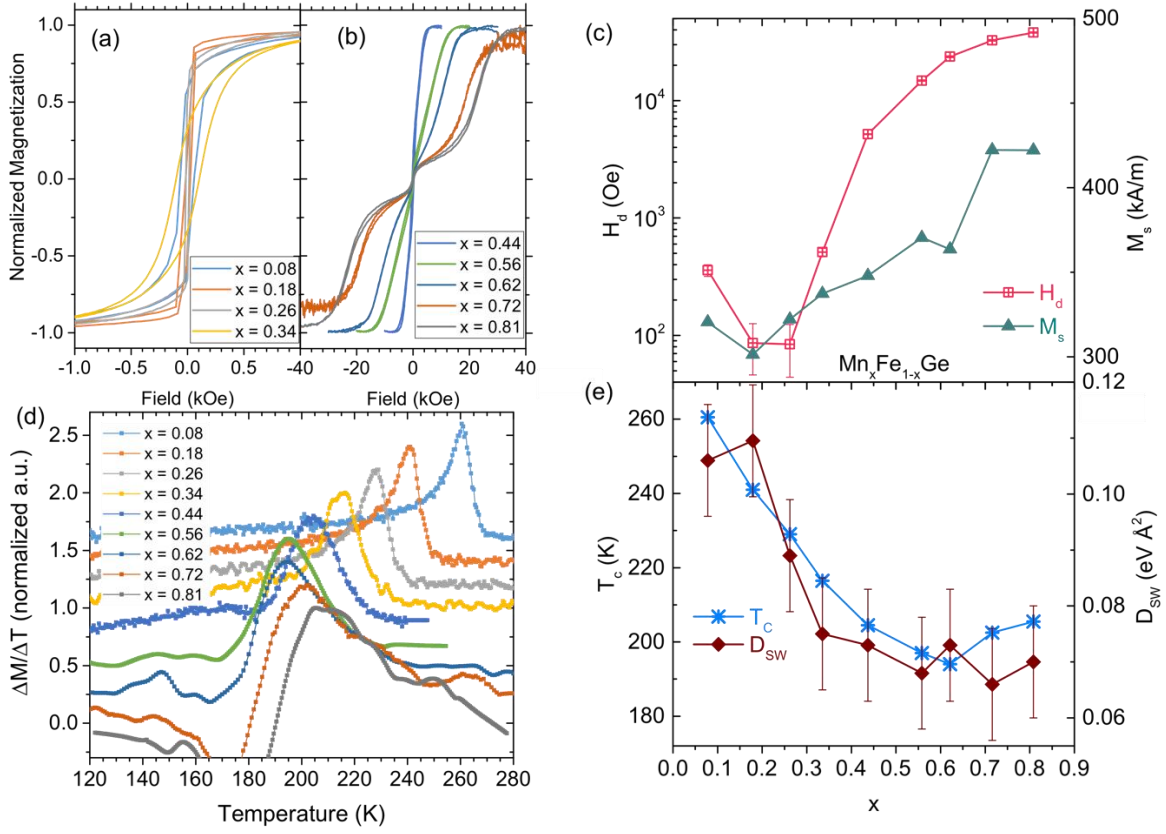

**FIG. 2.** Magnetometry characterization of $Mn_xFe_{1-x}Ge$ films. (a) and (b) show normalized magnetizations of films. (c) shows the saturation fields and magnetizations vs. $x$. While $M_s$ shows a relatively slow increase with higher manganese fractions, $H_d$ shows a dramatic increase due to the large DMI. (d) shows the temperature-dependent magnetization measurements under a small magnetic field (see the text). From the peak locations of $\Delta M/\Delta T$, we identified the critical temperatures $T_c$. (e) shows the trend of $T_c$ and spinwave stiffness constant, $D_{sw}$, as a function of $x$. $T_c$'s are extracted from (d) and $D_{sw}$ are extracted from Bloch-$T^{3/2}$ fittings of magnetization measurements at saturated fields (see supplemental material for individual fits).

**TABLE I.** Physical and magnetic properties of $Mn_xFe_{1-x}Ge$ films. $M_s$, $A$, and DMI are essential for the correct representation in micromagnetic simulations to account for the helical resonances.

| Sample # | Mn % | $\varepsilon_\perp$ % | $\varepsilon_{//}$ % | Ms kA/m | Hd Oe | λ nm | A  J/m $10^{-12}$ | DMI mJ/m$^2$ |
|---|---|---|---|---|---|---|---|---|
| 1 | 7.8 | -0.22 | 0.6 | 320 | 359 | 100.3 | 1.47 | 0.184 |
| 2 | 17.9 | -0.14 | 0.34 | 301 | 86 | 208.3 | 1.43 | 0.086 |
| 3 | 26.2 | -0.17 | 0.46 | 322 | 84 | 190.0 | 1.24 | 0.082 |
| 4 | 33.5 | -0.08 | 0.29 | 337 | 512 | 70.6 | 1.09 | 0.194 |
| 5 | 43.7 | -0.06 | 0.35 | 347 | 5172 | 21.9 | 1.09 | 0.625 |
| 6 | 55.8 | -0.22 | 0.63 | 370 | 14800 | 12.3 | 1.06 | 1.076 |
| 7 | 62.1 | -0.23 | 0.6 | 363 | 23670 | 10.0 | 1.10 | 1.375 |
| 8 | 71.6 | -0.27 | 0.68 | 422 | 32600 | 8.0 | 1.13 | 1.760 |
| 9 | 80.8 | -0.21 | 0.52 | 422 | 38000 | 7.6 | 1.19 | 1.951 |

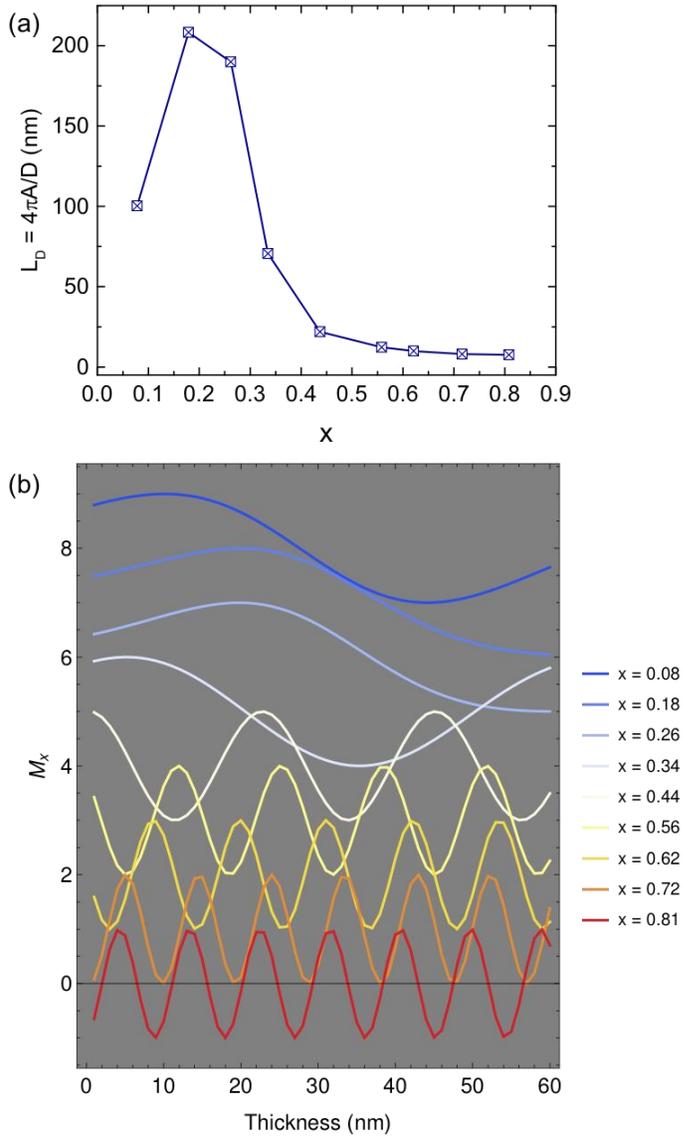

**FIG. 3.** (a) shows the helical period $L_D$ vs. $x$. (b) Equilibrium magnetization along the $x$-axis, $M_x$, that is parallel to the DC magnetic field, at 0 Oe field. Note a +1 offset is added to each successive curve for clarity. While the films with $x = 0.08$, 0.18, 0.26, and 0.34 manganese fractions present incomplete winding, the rest of the films have multiple windings.

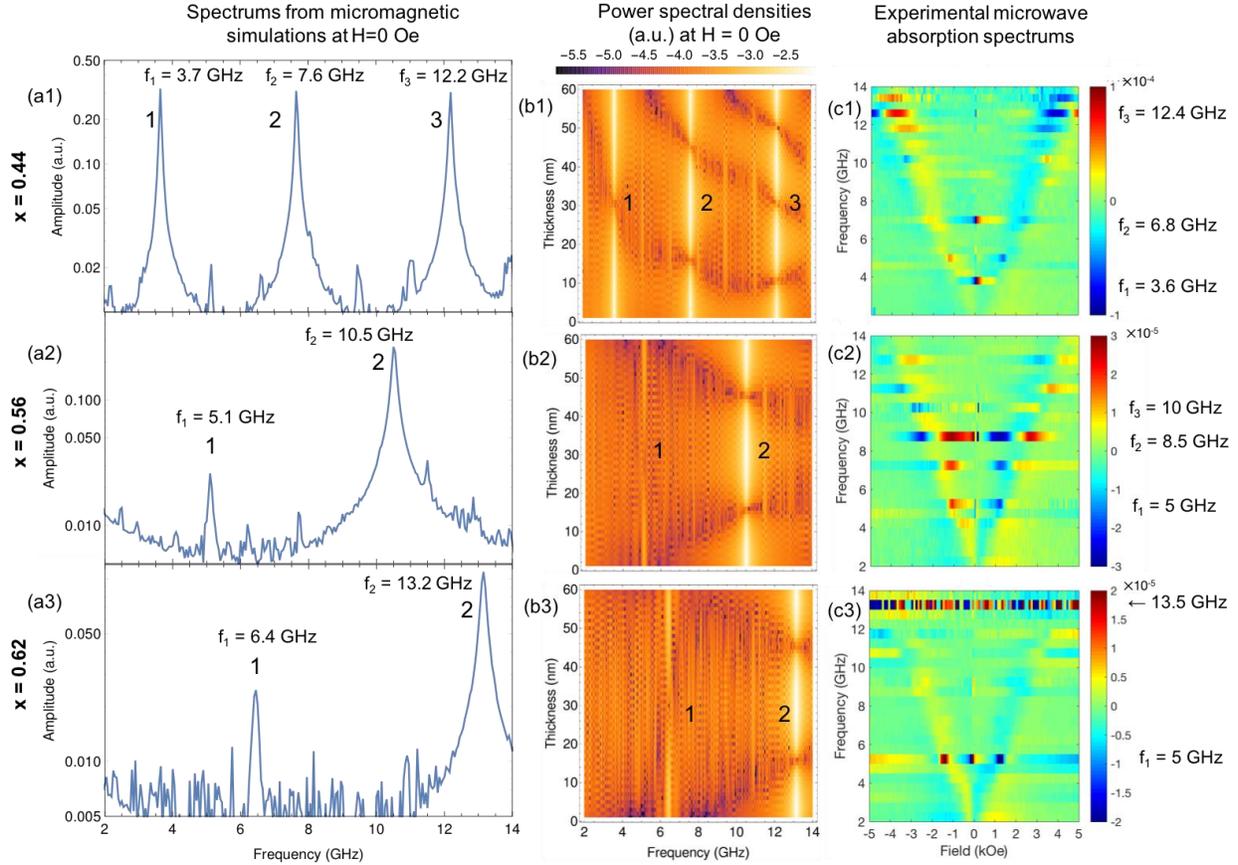

**FIG. 4.** Numerical and experimental results of microwave absorptions spectroscopy for the films with $x = 0.44$ (first row), $x = 0.56$ (second row), and $x = 0.62$ (third row) manganese fractions. (a1), (a2), and (a3) show the spinwave spectrum at $H = 0$ Oe field. (b1), (b2), and (b3) show the spatially resolved spinwave spectral densities as a function of thickness and frequency. We observe nodes and antinodes in the resonance profiles that is shaped by the helical structures. (c1), (c2), and (c3) show the experimental measurement of absorption spectral densities as a function of RF frequency and magnetic field. We note that we concentrate near-zero field resonance features to find the resonance frequencies of the helical phases.

# Supplemental Material for Engineering Dzyaloshinskii-Moriya interaction in B20 thin film chiral magnets

Emrah Turgut[1], Hanjong Paik[2], Kayla Nguyen[1], David A. Muller[1,3], Darrell G. Schlom[2,3], and Gregory D. Fuchs[1,3]

1. School of Applied and Engineering Physics, Cornell University, Ithaca, NY 14853, USA
2. Department of Materials Science and Engineering, Cornell University, Ithaca, NY 14853, USA
3. Kavli Institute at Cornell for Nanoscale Science, Ithaca, NY 14853, USA

In this supplemental material, we provide detailed information about X-Ray diffraction (XRD) analysis, lattice spacing calculations, magnetometry measurements, validation of Bloch-$T^{3/2}$ law for helimagnets, and micromagnetic simulation results.

## I. X-ray Diffraction Measurements

To accurately determine the volume of the film and simulate its micromagnetic behavior, we measure the thickness of our films by X-ray reflectivity (XRR). By fitting the XRR profiles of $Mn_{0.08}Fe_{0.92}Ge$, $Mn_{0.18}Fe_{0.82}Ge$, and $Mn_{0.62}Fe_{0.38}Ge$, we obtain the thicknesses of these three films to be approximately 60 nm, and we assume the rest of the films have the same thickness due to the same deposition rate.

We use *in-situ* RHEED to monitor the epitaxial growth of the B20 phase of $Mn_xFe_{1-x}Ge$ films. In Fig. S1(a), we show a RHEED pattern of the B20 crystalline phase of one of the films. The corresponding $\theta$–$2\theta$ XRD scan is shown in Fig. S1(b).

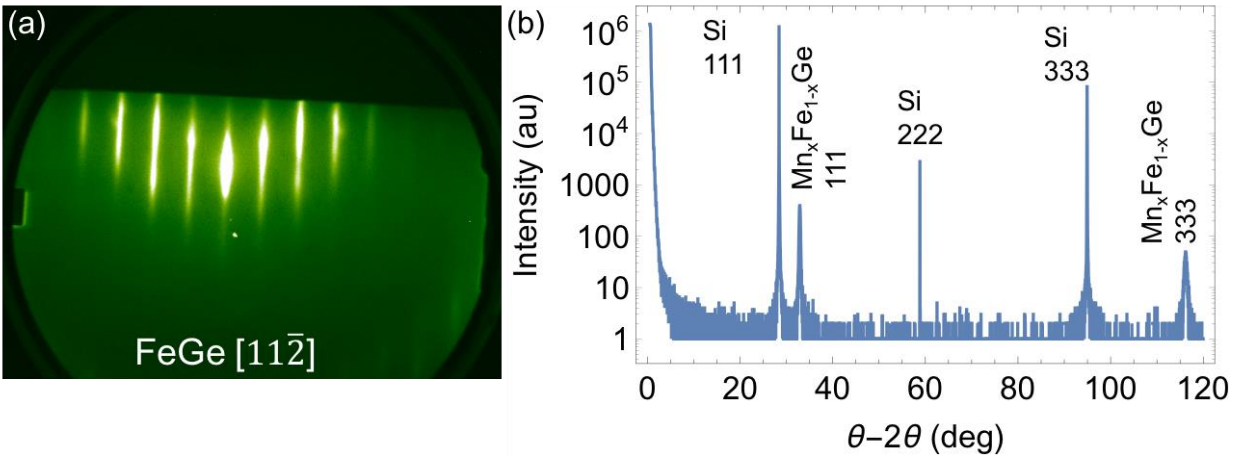

**Fig. S1 (a) RHEED image of $Mn_{0.18}Fe_{0.82}Ge$ film. (b) $\theta$–$2\theta$ XRD scan of $Mn_{0.08}Fe_{0.92}Ge$ film.**

To find the strains in the films, we measure the lattice $d$-spacing along the [111] and [100] directions, namely $d_{111}$ and $d_{100}$. These spacings are systematically calibrated against the known lattice constant of the silicon substrate. Then, we use the rhombohedral unit cell lattice formula [1]

$$\frac{1}{d_{hkl}} = \frac{1}{a}\left\{\frac{(1+\cos^2\alpha)(h^2+k^2+l^2)-\left(1-\tan^2\frac{\alpha}{2}\right)(hk+kl+lh)}{1+\cos\alpha-2\cos^2\alpha}\right\}^{1/2}, \qquad [S1]$$

and calculate the in-plane spacing of the films, which is then used to find the strains. To confirm the compositional fraction of the elements in each $Mn_xFe_{1-x}Ge$ film, we also perform energy-dispersive X-ray spectroscopy of at 12 keV and 18 keV electron energies, and use the average of the rates at these two energies as the Mn:Fe ratio of each film. By using these ratios and Vegard's law [S1], we find the expected bulk lattice constant for each film and show them with the measured ones in Fig. S2.

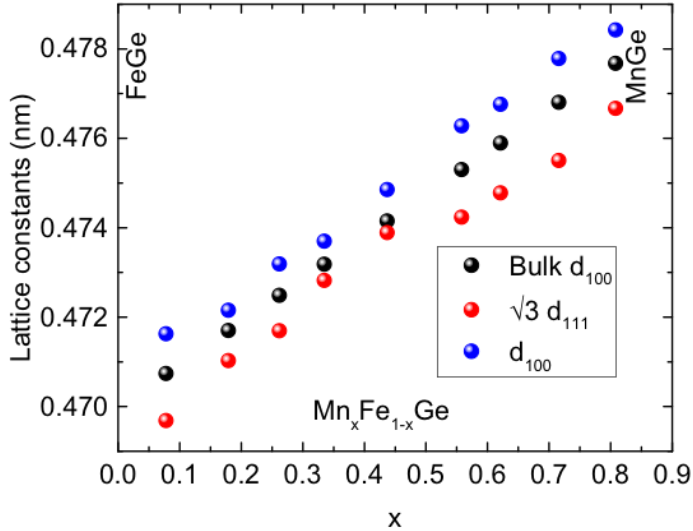

**Fig. S2. Measured at different direction and calculated expected lattice constants of $Mn_xFe_{1-x}Ge$ thin films.**

II. **Magnetometry measurements**

To find the saturation field and magnetization of the films, we perform magnetometry measurements using a Quantum Design VSM with an in-plane applied field. By finding the zero-crossings in the $\frac{d^2M}{dH^2}$ curve of each film as shown in Fig. S3, we obtain the saturation field $H_d$ as discussed in the main text. For the Mn$_{0.18}$Fe$_{0.82}$Ge film, the step size in magnetic field is coarse with respect to the saturation field, thus we use the $M$ vs. $H$ curve from Fig. 2(a) in the main text.

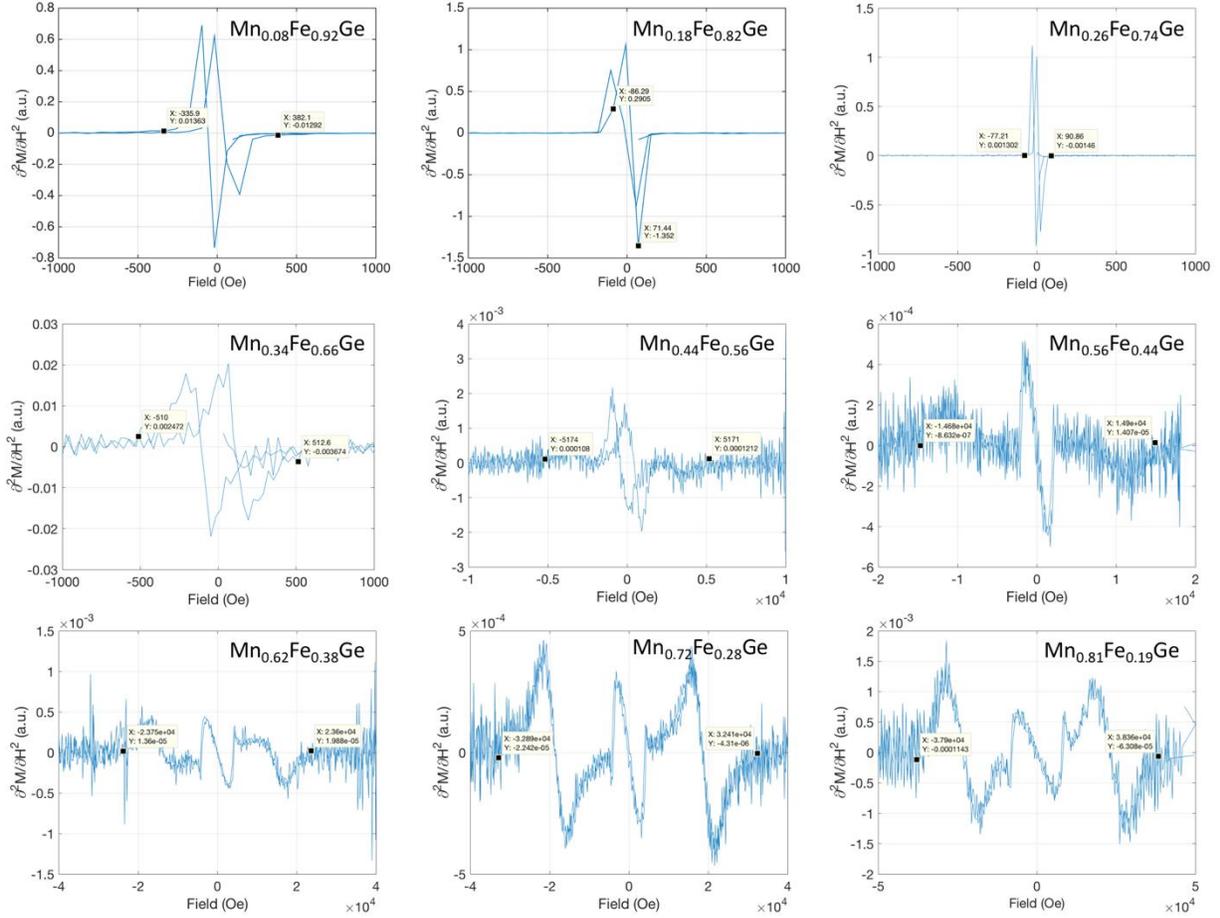

**Fig. S3.** $\frac{d^2M}{dH^2}$ **curves for each film, which is calculated by taking a numerical derivative of the $M$ with respect to $H$ measurements shown in Fig. 2 of the main text.**

To find spinwave stiffness constant $D_{SW}$, we apply a saturating large magnetic field and measure the temperature dependence of magnetization. We use the Bloch-$T^{3/2}$ law and fit the magnetization to Eq. (2) in the main text, obtaining $M_s$ and $D_{SW}$ at 0 K. In Fig. S4, we show these fits for each film. Then, we find $D_{SW}(T = 40$ K$)$ by using $D_{SW}(T) = D_{SW}(T = 0)[M_S(T)/M_S(T = 0)]$ [S2].

Although the Bloch-$T^{3/2}$ law has been widely used for ferromagnetic materials to obtain the symmetric exchange interaction constant, there are very few studies on the materials with DMI [S2]. Here we elaborate the validation of the Bloch-$T^{3/2}$ law for helimagnets as well.

The free energy of spinwaves for ferromagnetic materials is written as

$$H = -\frac{J}{2}\sum_j [S_j^+ S_{j+1}^- + S_j^- S_{j+1}^+] - J\sum_j S_j^z S_{j+1}^z, \quad [S2]$$

where $S_j$ represents a spin located at the $j$th site, $J$ is the symmetric exchange interaction, and $S_j^{\pm} = S_j^x \pm iS_j^y$ [S3]. With the additional DMI, this energy becomes

$$H = -\frac{\tilde{J}}{2}\sum_j [S_j^+ S_{j+1}^- + S_j^- S_{j+1}^+] - J\sum_j S_j^z S_{j+1}^z, \quad [S3]$$

where $\tilde{J} = |J + iD| = \sqrt{J^2 + D^2}$ [S4]. We note that the free energies in Eqs. (S2) and (S3) are atomistic versions, which are appropriate for the Bloch-$T^{3/2}$ law derivation. In contrast, we use the free energy of the micromagnetic version in the main text to have a coherent representation with micromagnetic simulations. The conversion between these two versions of the exchange terms is performed using the helical period [S4]:

| Helical period | Micromagnetic | Atomistic |
|---|---|---|
| $L_D$ | $4\pi A/D_m$ | $2\pi a_0 J/D_a$ |

where, $a_0$ is the lattice constant of the unit cell, $A$ is the micromagnetic exchange constant, $D_m$ and $D_a$ are the DMI of micromagnetic and atomistic energies, respectively. Thus, $D_a$ can be obtained by $D_m a_0/2$. By using the standard derivation of the Bloch law [S3], we find $D_{sw} \approx 2\tilde{J}Sa_0^2 = \frac{2g_e\mu_B A}{M_s}\left(1 + \frac{1}{2}Q^2 a_0^2\right) = \frac{g_e\mu_B H_D}{Q^2}\left(1 + \frac{1}{2}Q^2 a_0^2\right)$. If the helical period $L_D$ is much larger than the lattice constant $a_0$, the $D_{sw} = \frac{g_e\mu_B H_D}{Q^2}$ approximation can be used, which has been true for studies in B20 MnSi and FeGe [S5–S7]. For our films with high manganese fraction, however, this approximation results in a 4% error in the helical period.

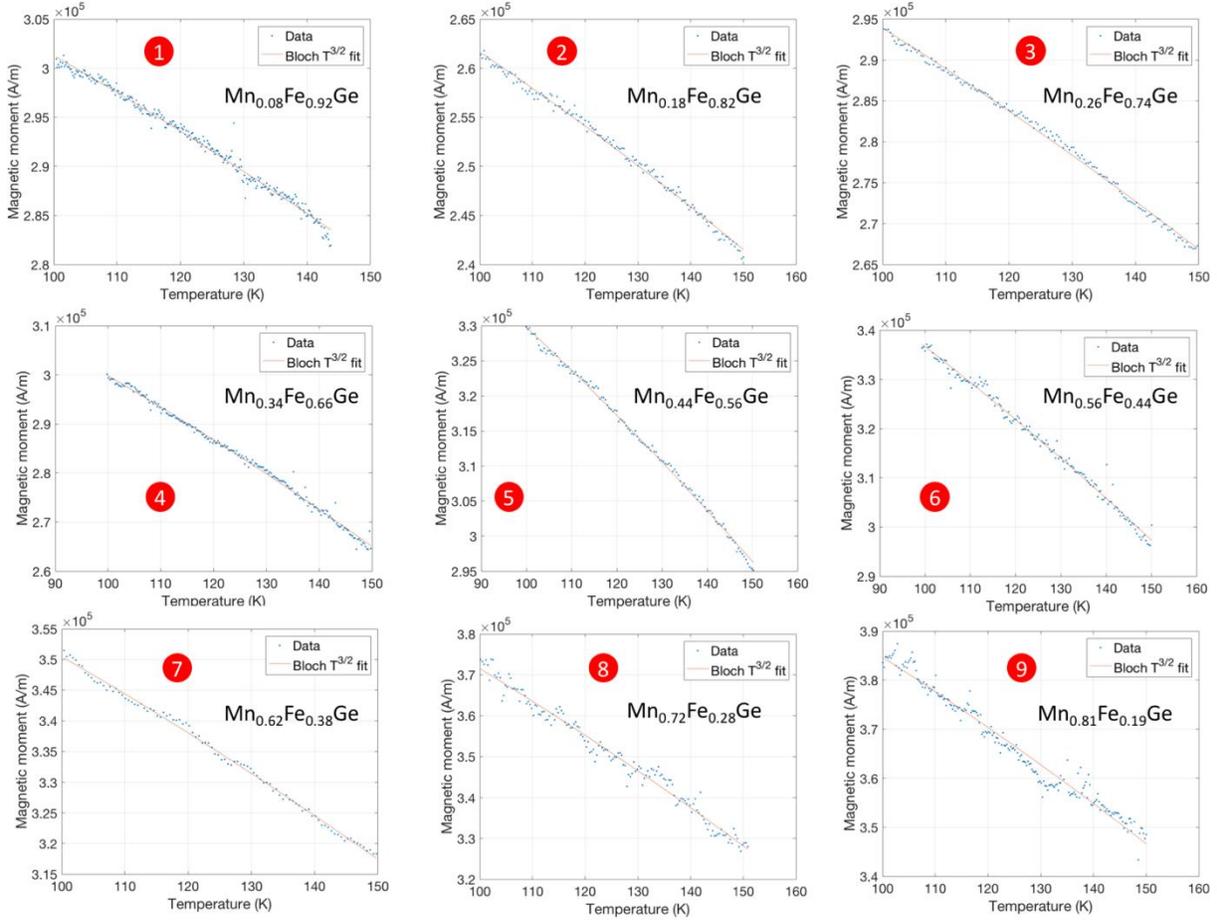

**Fig. S4.** *M* vs. *T* measurements at a saturating magnetic field and Bloch-$T^{3/2}$ fits.

III. **Micromagnetic simulations**

In order to find the resonance frequencies, we use the ringdown method, in which we apply a magnetic impulse after the ground spin-state is reached. This magnetic pulse creates oscillations and precession in the spin system. By calculating the spatially-resolved Fourier transform of these oscillations, we find the resonance frequencies and oscillation profiles (spinwave modes) along the film thickness. In Fig. S5, we show the thickness profiles of the natural oscillations at $H = 0$ Oe field.

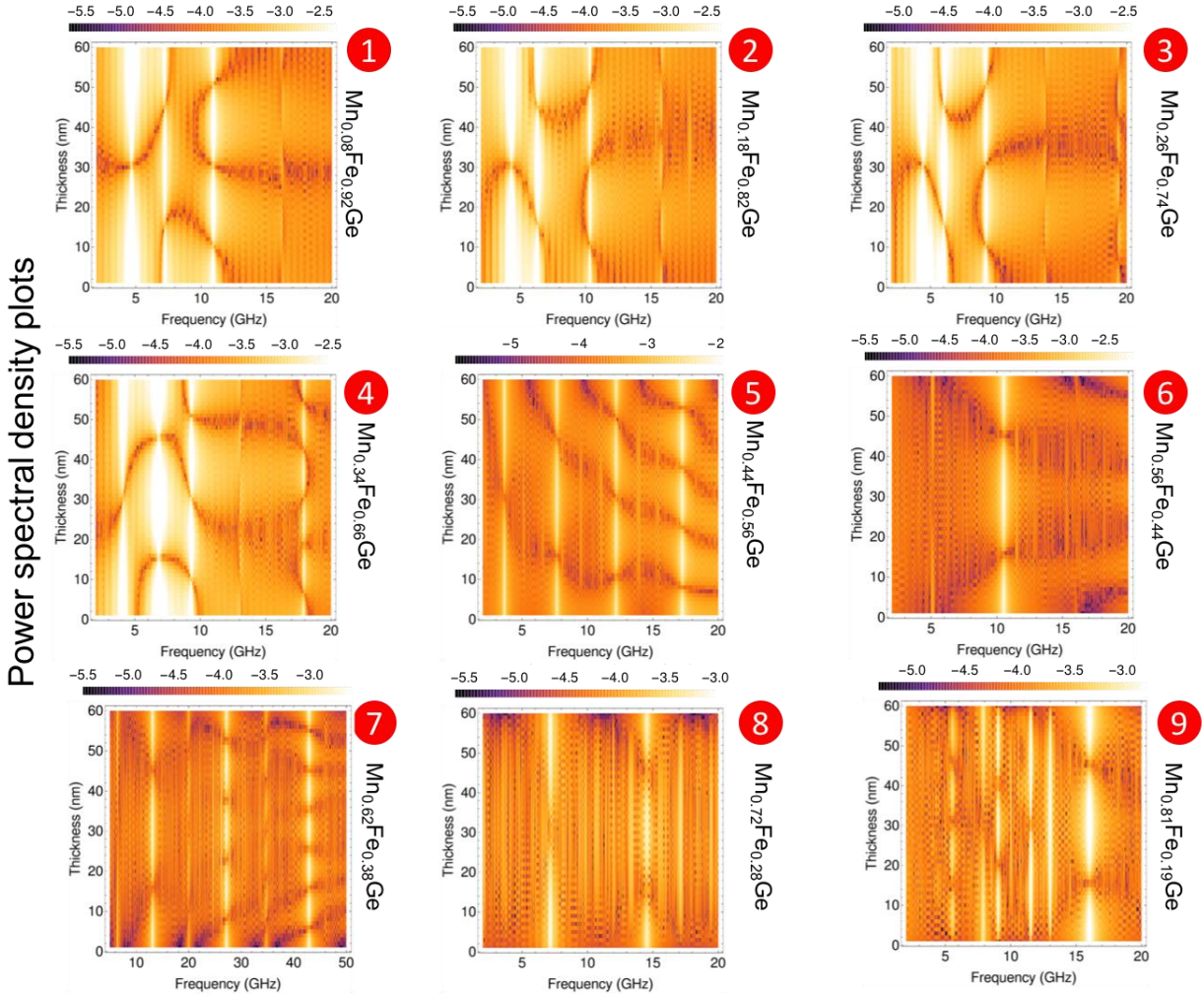

**Fig. S5. Thickness profiles of the natural oscillations obtained from ringdown micromagnetic simulations. Sample 7 is shown until 50 GHz to illustrate the scaling of the number of nodes and anti-nodes.**